\documentclass[reprint,amsmath,amssymb,aps,nofootinbib,superscriptaddress]{revtex4-2}

\usepackage[utf8]{inputenc}
\usepackage{graphicx}
\usepackage{xcolor}
\usepackage{amsmath,amsfonts,latexsym,amssymb}
\usepackage{bm}
\usepackage{bbm}
\usepackage{dsfont}
\usepackage{siunitx}
\usepackage[colorlinks=true,allcolors=blue]{hyperref}

\newcommand{\beq}{\begin{equation}}
\newcommand{\eeq}{\end{equation}}
                  
\newcommand{\benum}{\begin{enumerate}}
\newcommand{\eenum}{\end{enumerate}}
                    
\newcommand{\bit}{\begin{itemize}}
\newcommand{\eit}{\end{itemize}}
\newcommand{\xhat}{\hat{\T{x}}}
\newcommand{\yhat}{\hat{\T{y}}}
\newcommand{\zhat}{\hat{\T{z}}}

\newcommand{\bea}{\begin{eqnarray}}
\newcommand{\eea}{\end{eqnarray}}

\newcommand{\bev}{\begin{verbatim}}
\newcommand{\eev}{\end{verbatim}}

\newcommand{\T}[1]{\textbf{#1}}
\newcommand{\I}[1]{\textit{#1}}

\newcommand{\zl}[1]{\label{eqn:#1}}
\newcommand{\zr}[1]{(\ref{eqn:#1})} 
\newcommand{\zfl}[1]{\protect\label{fig:#1}}
\newcommand{\zfr}[1]{\ref{fig:#1}} 

\newcommand{\ba}{\left\{ \begin{array}{lr}}
\newcommand{\ea}{\end{array}\right.}

\newcommand{\blist}[1]{
 \begin{list}{#1}
 \begin{align}
	 arrow
 \end{align}
 $\checkmark\star
  { \setlength{\itemsep}{3pt}
     \setlength{\parsep}{2pt}
     \setlength{\topsep}{3pt}
     \setlength{\partopsep}{0pt}
     \setlength{\leftmargin}{1em}
     \setlength{\labelwidth}{1em}
     \setlength{\labelsep}{0.5em} } }
\newcommand{\elist}{
  \end{list}  }

\DeclareMathSymbol{\vartheta}{\mathalpha}{letters}{"12}
\DeclareMathSymbol{\theta}{\mathalpha}{letters}{"23}
\DeclareMathSymbol{\phi}{\mathalpha}{letters}{"27}
\DeclareMathSymbol{\varphi}{\mathalpha}{letters}{"1E}

\newcommand{\bef}
{
\begin{figure}[htbp]
\centering
}

\newcommand{\eef}{\end{figure}}

\newcommand{\affA}{Department of Chemistry, University of California, Berkeley, Berkeley, CA 94720, USA.}

\begin{document}

\title{Prethermal rotating-frame solid echo in a dipolar nuclear-spin network}

\author{Quentin Reynard-Feytis}\affiliation{\affA}
\author{William Beatrez}\affiliation{\affA}
\author{Leo Joon Il Moon}\affiliation{\affA}
\author{Emanuel Druga}\affiliation{\affA}
\author{Ashok Ajoy}\affiliation{\affA}
    
\begin{abstract}
Floquet prethermalization can endow interacting quantum solids with long-lived, approximately conserved quantities, enabling Hamiltonian engineering and new dynamical probes. Using a hyperpolarized network of dipolar-coupled $^{13}$C nuclear spins in diamond driven by pulsed spin-locking, we access a rotating-frame prethermal plateau with quasi-conserved transverse magnetization and cycle-resolved inductive readout. Within this prethermal manifold we observe a robust \emph{rotating-frame solid echo}: after an apparent decay of the rotating-frame free-induction signal over a delay $\tau$, the magnetization revives at time $2\tau$ following a single $(\alpha)_y$ pulse, with maximum amplitude near $\alpha\simeq\pi/2$. The echo envelope decays as a stretched exponential with characteristic time $T_2'\approx 13\,$ms. Analytical arguments and toy-model simulations attribute the revival to Floquet micromotion that transfers coherences between operator subspaces, so that only a subset of the many-body dephasing dynamics is inverted by the $y$ pulse. These results translate classic echo physics into the prethermal rotating frame and point to continuously interrogated prethermal spin ensembles as a versatile platform for high-throughput spectroscopy, Hamiltonian engineering, and long-duration quantum sensing.
\end{abstract}

\maketitle

\begin{figure}[t]
    \centering
    \includegraphics[width=\columnwidth]{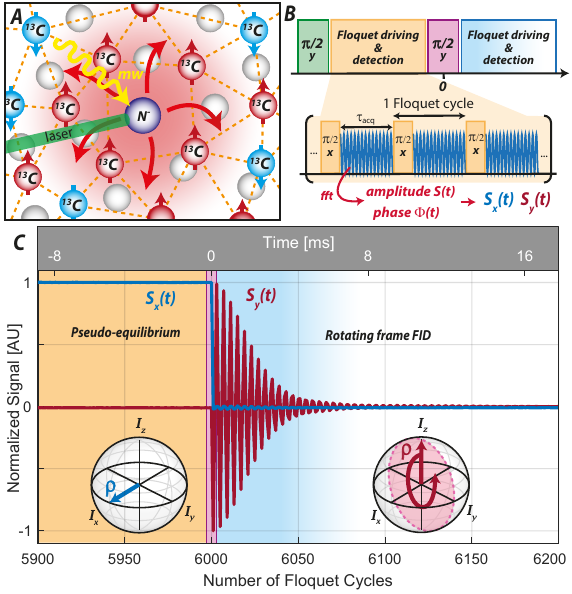}
    \caption{\T{Rotating-frame FID under Floquet prethermalization.}
(A) \I{Platform.} Hyperpolarized $^{13}$C nuclei in diamond form a dilute 3D dipolar network (dashed couplings) via optically pumped NV centers (yellow).
(B) \I{Floquet drive and readout.} A $(\pi/2)_x$ pulse prepares transverse magnetization, followed by a pulsed spin-locking train that both drives the system and enables cycle-resolved sampling of the Larmor precession between pulses; a subsequent $(\pi/2)_y$ pulse tips the prethermal magnetization to initiate the rotating-frame FID.
(C) \I{Dynamics.} $S_x(t)$ exhibits a long-lived prethermal plateau (yellow), while $S_y(t)$ shows a rapidly dephasing oscillatory response after the tip pulse at $t{=}0$. Upper axis: time; lower axis: Floquet cycles.}
    \zfl{fig1}
\end{figure}


\emph{Introduction---}Periodic (Floquet) driving provides a powerful route to engineer effective Hamiltonians and stabilize nonequilibrium phases~\cite{goldman2014periodically,bukov2015universal}, but generic interacting systems ultimately absorb energy from the drive and heat toward featureless states. When the drive is fast, heating can be parametrically slow and the dynamics can become trapped for long times in a \emph{prethermal plateau} governed by an emergent effective Hamiltonian and approximate conservation laws~\cite{dalessio2014long,lazarides2014equilibrium,abanin2015exponentially,bukov2015prethermal,mori2016rigorous,weidinger2017floquet,else2017prethermal,machado2020longrange,mori2018thermalization,ho2023quantum,rubio2020floquet}. Recent experiments have used this idea to extend coherence and enable long-time control in dense spin ensembles~\cite{beatrez2021floquet,peng2021floquet}. Related long-lived quasi-stationary states under periodic pulse trains were already identified in solid-state NMR, in the language of spin thermodynamics and quasi-equilibria~\cite{maricq1987spin,maricq1990long,sakellariou1998quasi}.
\newline\indent
In our platform---a dipolar network of $^{13}$C nuclei in diamond---pulsed spin-locking acts as a Floquet drive that generates a long-lived prethermal state with quasi-conserved transverse magnetization~\cite{beatrez2021floquet}. Combined with optical hyperpolarization, the resulting signal-to-noise permits cycle-resolved inductive detection of the rotating-frame magnetization without averaging. This quasi-continuous readout effectively promotes rotating-frame dynamics to an experimentally accessible ``laboratory frame'' in which one can directly track the driven many-body evolution.
\newline\indent
Here we report a surprising consequence of this prethermal rotating-frame control: a simple pulse sequence produces a \emph{solid-echo-like revival} even though the underlying dynamics arise from many-body dipolar interactions rather than static field inhomogeneity. In conventional NMR, the solid echo can refocus dephasing in small coupled-spin systems but becomes imperfect in extended dipolar networks~\cite{solomon1958multiple,powles1962double,mansfield1965multiple,waugh1967multiple,li2007generating}. We find that within the prethermal plateau, a single $(\alpha)_y$ ``echo'' pulse applied after a delay $\tau$ generates a pronounced revival at $2\tau$, with maximal amplitude near $\alpha\approx\pi/2$ and a stretched-exponential decay versus $\tau$. We interpret the effect as a Floquet analogue of the classic solid echo: micromotion under the drive periodically transfers the density matrix between operator subspaces, so that only the components coupled to $I_y$ are inverted and refocused by the $y$ pulse.

\par\indent
Echo phenomena are typically characterized by stepping a delay $\tau$ and recording a signal during a separate acquisition window, yielding a nominally two-dimensional experiment~\cite{ernst1987principles,freeman1998spin}. A key advantage of working in the prethermal rotating frame is that the observable Larmor precession can be sampled quasi-continuously between drive pulses, so that the echo physics is accessed with a large reduction in experimental overhead. Below, we first summarize the rotating-frame free induction decay (FID) and prethermal plateau in our driven dipolar solid (Fig.~\zfr{fig1}), then present the rotating-frame solid-echo protocol and observations (Fig.~\zfr{fig2}), and finally develop an intuitive mechanism supported by toy-model simulations (Figs.~\zfr{fig3} and~\zfr{fig4}).


\emph{Experimental platform and rotating-frame FID---}Our system comprises a three-dimensional, randomly positioned lattice of $^{13}$C nuclear spins in diamond at natural abundance (1.1\%), with a spin density of $\approx 1$ spin per nm$^3$~\cite{ajoy2019hyperpolarized}. Spins are optically hyperpolarized via NV centers~\cite{ajoy2018enhanced,ajoy2018orientation,pillai2023electron} and shuttled to high field for inductive readout and control~\cite{beatrez2021floquet}. The spins interact through the secular dipolar Hamiltonian $\mathcal{H}_{zz}$,
\begin{equation}
  \mathcal{H}_{zz}=\sum_{i<j} d_{ij}\bigl(3I_{iz}I_{jz}-\mathbf{I}_i\!\cdot\!\mathbf{I}_j\bigr),
  \label{ham1}
\end{equation}
with median coupling $\langle d_{ij}\rangle\equiv J\simeq\SI{660}{Hz}$.
\newline\indent
The control protocol, Fig.~\zfr{fig1}B, uses pulsed spin-locking~\cite{rhim1976multiple}: $(\theta)_x$ pulses of duration $t_p=50.5\,\mu$s separated by free-evolution windows $\tau=88.5\,\mu$s, forming a Floquet cycle of period $\tau_{\rm FC}=t_p+\tau=109\,\mu$s. Between pulses, the Larmor precession is sampled and its amplitude and phase extracted, yielding rotating-frame quadratures $S_x=\langle I_x(t)\rangle$ and $S_y=\langle I_y(t)\rangle$~\cite{sahin2022continuously, moon2025high}.
\newline\indent
For rapid pulsing ($J\tau_{\rm FC}\ll1$) and $\theta\not\approx\pi$, the drive is well described by the leading Magnus term~\cite{mananga2011introduction,magnus1954exponential,wilcox1967exponential,blanes2009magnus},
\begin{equation}
  \Bar{\mathcal{H}}_{F}=-\frac{1}{2}\sum_{i<j} d_{ij}\bigl(3I_{ix}I_{jx}-\mathbf{I}_i\!\cdot\!\mathbf{I}_j\bigr),
  \zl{avgHam}
\end{equation}
which approximately conserves $I_x$~\cite{haeberlen1976high}. After a transient, the system enters a Floquet prethermal plateau~\cite{maricq1987spin,maricq1990long,sakellariou1998quasi,linden2009quantum,weidinger2017floquet,mori2018thermalization,ho2023quantum,rubio2020floquet,saha2023cascaded} that appears as a long-lived $S_x(t)$ signal (yellow, Fig.~\zfr{fig1}C) and can persist for $T_2'\gg T_2^*$~\cite{beatrez2021floquet}. In previous work we observed a plateau lifetime $T_2'\approx 90\,$s, over four orders of magnitude longer than the conventional FID decay time~\cite{beatrez2021floquet}.
\newline\indent
Fig.~\zfr{fig1}B also shows the effect of tipping the spins away from the $\xhat$ axis with a $(\pi/2)_y$ pulse (purple block). This produces a rapidly dephasing oscillatory response in $S_y(t)$ (blue region in Fig.~\zfr{fig1}C), corresponding to precession and decay in the $\yhat$--$\zhat$ plane. The decay occurs on the scale of $\simeq 10\,$ms (\,\,$\sim$20 Floquet cycles). Overall, the dynamics in Fig.~\zfr{fig1}C can be viewed \emph{loosely} as a rotating-frame analogue of the conventional lab-frame FID, with free evolution replaced by evolution under the Floquet drive and observation windows interspersed throughout. We emphasize, however, that this analogy is not exact: in a conventional FID the observed $T_2^*$ decay reflects static-field inhomogeneity and coupling to paramagnetic defects, in addition to dipolar dephasing, whereas the pulsed spin-lock largely averages out dephasing from resonance offsets and slowly varying local fields (e.g., paramagnetic-defect fields and $B_0$ inhomogeneity). In the present setting the remaining decay is therefore dominated by dipolar couplings (with residual contributions from pulse imperfections), making the ``rotating-frame FID'' language a useful approximation for interpreting the subsequent echo dynamics.


\begin{figure}[t]
    \centering
    \includegraphics[width=\columnwidth]{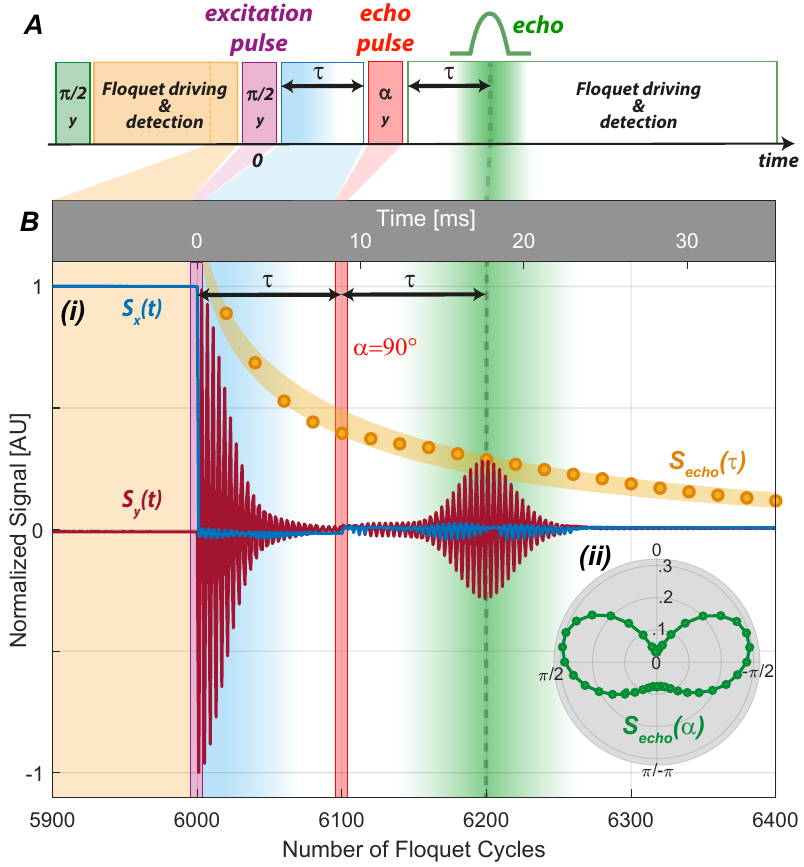}
    \caption{\T{Rotating-frame solid echo.}
(A) \I{Protocol.} After the tip pulse, the rotating-frame FID dephases for a delay $\tau$; a single echo pulse $(\alpha)_y$ then produces a revival at $2\tau$ during continued Floquet driving and quasi-continuous detection.
(B) \I{Echo formation and decay.} Single-shot traces $S_x(t)$ (blue) and $S_y(t)$ (red) for $\alpha=90^\circ$ applied at $\tau=8.85$ ms (100 Floquet cycles, shaded pink). The peak echo amplitude $S_{\rm echo}$ (yellow points) decays with $\tau$ as a stretched exponential, yielding $T_2'=13.4\pm1.3$ ms (shaded region: 95\% confidence interval). Inset: $S_{\rm echo}(\alpha)$ is maximal near $\alpha\simeq\pi/2$, analogous to the conventional solid echo.}
    \zfl{fig2}
\end{figure}

\emph{Rotating-frame solid echo---}We now explore the possibility of exciting phenomena similar to a conventional solid echo~\cite{powles1962double,mansfield1965multiple,solomon1958multiple,waugh1967multiple,powles1962measurement,mansfield1971pulsed,freeman1998spin}, but in the prethermal rotating frame. The sequence is illustrated in Fig.~\zfr{fig2}A, with the experimentally measured $S_x(t)$ and $S_y(t)$ components shown in Fig.~\zfr{fig2}B.
\newline\indent
Initially, the spins prethermalize along the $\xhat$ axis, producing the plateau in $S_x(t)$ (blue trace before $t=0$ in Fig.~\zfr{fig2}B). We then apply a pulse (purple in Fig.~\zfr{fig2}A) and take $t=0$ to coincide with this point. The spins exhibit a rotating-frame FID and the signal nearly fully decays after evolving under the Floquet Hamiltonian for a duration $\tau$. Following this, we apply an $(\alpha)_y$ pulse (red box), referred to as the echo pulse.
\newline\indent
Remarkably, after an additional delay $\tau$, an echo forms (green shaded region in Fig.~\zfr{fig2}B). The revival is prominent in $S_y(t)$ (red trace), while $S_x(t)$ (blue trace) remains near zero. The echo is observed across a range of $\alpha$ angles; the variation of the echo amplitude with $\alpha$ is shown in the polar inset Fig.~\zfr{fig2}B(ii). The strongest echo occurs close to (but not exactly) $\alpha=\pi/2$, mirroring the conventional solid echo where the maximal echo intensity follows a $\pi/2$ pulse~\cite{powles1962double,mansfield1965multiple,mansfield1971pulsed}. Experimentally we observe a small offset ($\approx 5^\circ$), plausibly arising from a slight resonance offset during the pulses.
\newline\indent
The yellow points in Fig.~\zfr{fig2}B show the maximum echo amplitude $S_{\rm echo}$ as a function of $\tau$. The envelope follows $S_{\rm echo}(\tau)=\exp\!\left[-\left(\tau/T_2'\right)^{\eta}\right]$ with stretching factor $\eta\approx 1/2$ and fitted decay time $T_2'=13.4\pm 1.3\,$ms. Similar stretched-exponential behavior is common in spin-lock and echo relaxometry in solids~\cite{vega1985relaxation, henrichs1984nuclear}. In the same vein as Fig.~\zfr{fig1}C, the echo in Fig.~\zfr{fig2}B can be viewed as a rotating-frame analogue of the conventional lab-frame solid echo.


\begin{figure*}[t]
    \centering
    \includegraphics[width=\textwidth]{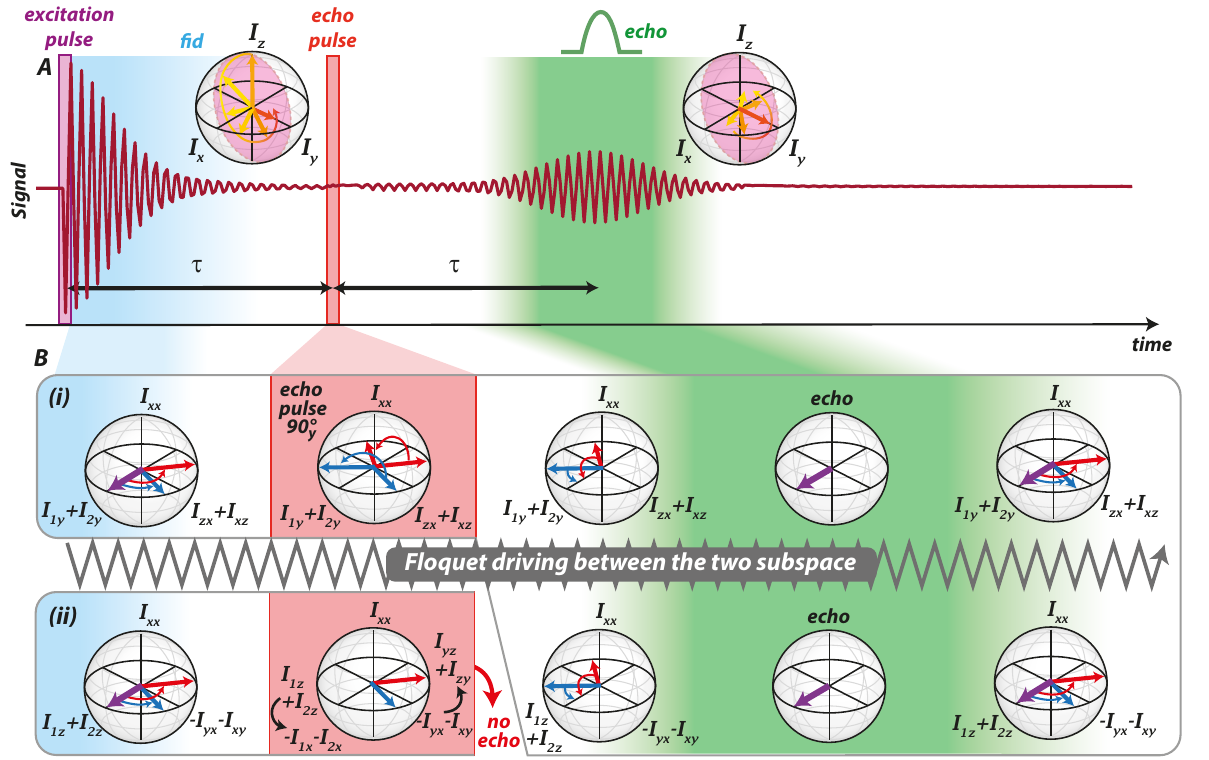}
    \caption{\T{Micromotion-assisted refocusing mechanism.}
(A) Pulse sequence (shaded) overlaid with the experimental $S_y(t)$ trace from Fig.~\zfr{fig2}B, highlighting dephasing during the rotating-frame FID (blue) and revival at $2\tau$ (green).
(B) Minimal two-spin picture. Under the zeroth-order average Hamiltonian $\propto I_{xx}$, evolution alternates between two operator subspaces (i) and (ii) that are coupled by Floquet micromotion (grey). The $(\pi/2)_y$ pulse inverts only the $I_y$-coupled subspace, refocusing a subset of coherences into an echo while the complementary subspace does not refocus.}
    \zfl{fig3}
\end{figure*}

\emph{Mechanism---}To investigate the mechanism underlying the rotating-frame solid echo, we turn to analytical models and simulations (Figs.~\zfr{fig3} and~\zfr{fig4}). Fig.~\zfr{fig3}A shows the experimental $S_y(t)$ trace and schematically depicts the rotating-frame FID (blue region) and echo formation (green region). During the FID, the magnetization evolves primarily in the $\xhat$--$\zhat$ plane, leading to near-complete decay; the subsequent echo reflects a partial refocusing of this dephasing.
\newline\indent
Fig.~\zfr{fig3}B provides a diagrammatic explanation using a minimal model of two dipolar-coupled spins~\cite{wells1965nmr,bain1978modulation}. The Floquet drive generates an average Hamiltonian $\Bar{\mathcal{H}}_{F}$, Eq.~\zr{avgHam}. To illustrate dephasing and refocusing, we consider two independent spin pairs with different effective couplings and represent the dynamics on two operator ``Bloch spheres'' with axes $\{I_{1y}{+}I_{2y},\, I_{zx}{+}I_{xz},\, I_{xx}\}$ and $\{I_{1z}{+}I_{2z},\, -I_{yx}{-}I_{xy},\, I_{xx}\}$ [Fig.~\zfr{fig3}B(i) and (ii)]. Here we use bilinear operators $I_{\mu\nu}=2I_{1\mu}I_{2\nu}$. Because the $\mathbf{I}_i\!\cdot\!\mathbf{I}_j$ part of $\Bar{\mathcal{H}}_{F}$ commutes with the relevant collective components, the observed decay is governed by the $I_{xx}$ part.
\newline\indent
Starting with an initial $I_{1y}{+}I_{2y}$ component (purple arrows), evolution under $I_{xx}$ yields
\begin{equation}
\begin{split}
    I_{1y}{+}I_{2y}\xrightarrow[\tau]{\,d'_{12}I_{xx}\,}\cos(\alpha_{12})\,(I_{1y}{+}I_{2y})\\
    +\sin(\alpha_{12})\,(\overbrace{2I_{1z}I_{2x}}^{I_{zx}}+\overbrace{2I_{1x}I_{2z}}^{I_{xz}}),
\end{split}
    \label{ev1}
\end{equation}
where $d'_{ij} = -\frac{3}{4}d_{ij}$,  and $\alpha_{12}=d'_{12}t$ is the evolution angle within this subspace. Different couplings produce ``fast'' and ``slow'' components (red and blue arrows), whose relative dephasing drives the FID-like decay, analogous to the conventional solid echo~\cite{powles1962double,mansfield1965multiple}.
\newline\indent
Similarly, an initial $I_{1z}{+}I_{2z}$ component evolves as
\begin{equation}
\begin{split}
    I_{1z}{+}I_{2z}\xrightarrow[\tau]{\,d'_{12}I_{xx}\,}\cos(\alpha_{12})\,(I_{1z}{+}I_{2z})\\
    -\sin(\alpha_{12})\,(I_{yx}{+}I_{xy}),
\end{split}
    \label{ev2}
\end{equation}
leading to dephasing in the complementary subspace [Fig.~\zfr{fig3}B(ii)]. Importantly, the Floquet micromotion continuously transfers amplitude between the two subspaces (grey oscillation in Fig.~\zfr{fig3}B), so dephasing proceeds even though the system is repeatedly ``kicked'' between manifolds.
\newline\indent
The action of the $(\pi/2)_y$ echo pulse is shown in Fig.~\zfr{fig3}B(i): it reflects vectors in the $I_y$-coupled subspace, swapping the fast and slow components. Subsequent evolution for an additional time $\tau$ refocuses these components and forms an echo (thick purple arrow). In contrast, vectors residing in the $\{I_{1z}{+}I_{2z},-I_{yx}{-}I_{xy},I_{xx}\}$ subspace are not inverted into an echo-producing configuration by a $y$ pulse [Fig.~\zfr{fig3}B(ii)], so they do not refocus. In practice, leakage between subspaces induced by higher-order Floquet terms, many-body interactions, and pulse imperfections naturally limits the echo amplitude.

\begin{figure*}[t]
    \centering
    \includegraphics[width=\textwidth,height=0.33\textheight,keepaspectratio]{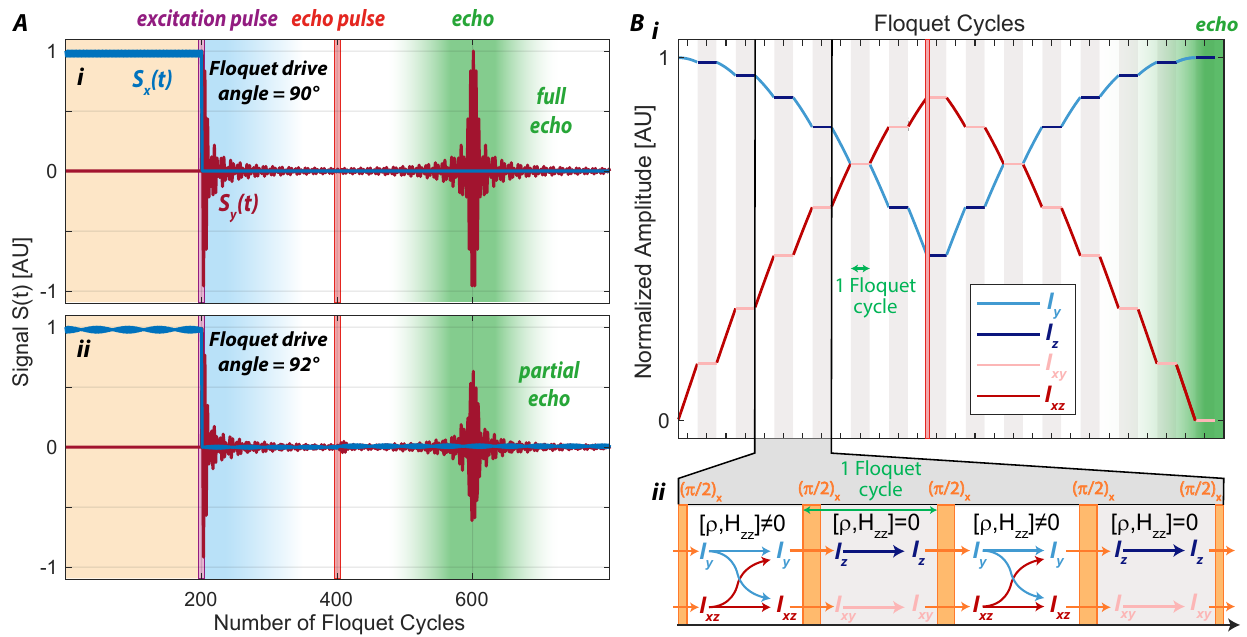}
    \caption{\T{Toy-model simulations reproduce the rotating-frame solid echo.}
(A) Simulated quadratures $S_x(t)$ (blue) and $S_y(t)$ (red) for an ensemble of 60 dipolar-coupled $^{13}$C spin pairs with uniformly distributed couplings ($\pm5$ kHz), using the experimental cycle timing and an $\alpha=90^\circ$ echo pulse. Ideal $90^\circ$ drive pulses yield a strong symmetric revival at $2\tau$ (i), whereas a small drive-angle error ($92^\circ$) mixes operator subspaces and reduces the refocused fraction (ii).
(B) Representative density-matrix components for a single pair: alternating Floquet cycles correspond to commuting/noncommuting evolution (white/grey), and the echo pulse reverses the dephasing for the invertible components (zoom in, ii).}
    \zfl{fig4}
\end{figure*}

Fig.~\zfr{fig4} illustrates numerical simulations that reproduce the echo formation. For simplicity, we simulate 60 independent $^{13}$C spin pairs coupled by dipolar interactions with coupling constants uniformly distributed over $\pm 5$ kHz, using similar cycle timing as the experiment ($\tau_{\rm acq}=38\,\mu$s and $\tau$ spanning 200 Floquet cycles). We assume a perfect $\alpha=90^{\circ}$ echo pulse.
\newline\indent
Panels Fig.~\zfr{fig4}A(i) show that for ideal $90^{\circ}$ Floquet pulses in the \emph{two-spin} model, the micromotion is perfectly synchronized with the drive: over successive Floquet cycles the density matrix is coherently shuttled between an echo-active manifold (spanned by $I_{1y}{+}I_{2y}$ and $I_{xz}{+}I_{zx}$, where dephasing under the dipolar interaction occurs) and a commuting sector proportional to $I_{1z}{+}I_{2z}$ [Fig.~\zfr{fig4}B(ii)]. Because the echo pulse is applied stroboscopically (here after an even number of cycles), the state lies in the $I_y$-coupled manifold at the pulse time, so the $(\pi/2)_y$ pulse fully inverts the accrued dephasing and yields a strong, symmetric revival at $2\tau$ (maximal near $\alpha=\pi/2$). This idealized behavior highlights an important distinction between the two-body simulation and the experiment: in a many-spin network and with realistic pulse errors, the density matrix can leak into the non-refocusable components of the complementary manifold (e.g., $-I_{yx}-I_{xy}$) at the stroboscopic times, reducing the refocused fraction and producing a partial echo.
\newline\indent
Fig.~\zfr{fig4}A(ii) repeats the simulation with a small drive-angle error ($92^{\circ}$). The resulting micromotion mixes the density matrix across the two subspaces, so that only the portion residing in the $\{I_{1y}{+}I_{2y}, I_{zx}{+}I_{xz}, I_{xx}\}$ manifold can be refocused; this yields a reduced (partial) echo, consistent with the experimental dependence of $S_{\rm echo}$ on $\alpha$.
\newline\indent
To further illustrate the alternating commuting/noncommuting evolution underlying the echo, Fig.~\zfr{fig4}B tracks representative density-matrix components of a single spin pair over successive Floquet cycles. Even and odd cycles correspond to evolution in different subspaces (white/grey shading), producing periods of dephasing interleaved with intervals of approximate conservation. The echo pulse reverses the dephasing dynamics for the invertible components, resulting in an echo symmetric about the pulse time.

\emph{Outlook---}This experiment demonstrates the feasibility of designing NMR experiments in the rotating frame that closely mimic familiar lab-frame spin dynamics, while benefiting from quasi-continuous, cycle-resolved readout. Translating dynamics from the lab frame to the prethermal rotating frame opens several intriguing possibilities. First, the driven transverse magnetization can be tracked over long periods in the $\xhat$--$\yhat$ plane, enabling reconstruction of stable, large-excursion spin trajectories~\cite{sahin2022continuously}.
\newline\indent
Second, Floquet engineering provides a flexible way to dynamically modify the effective Hamiltonian, analogous to tailoring static Zeeman or dipolar interactions, creating a versatile toolbox for exploring and controlling many-body spin dynamics~\cite{ramanathan2024universal,gopalakrishnan2007quenching}.
\newline\indent
More broadly, long-lived driven prethermal order can also stabilize dynamical phases such as discrete time crystals in dipolar ensembles and related platforms~\cite{else2016floquet,else2020discrete,choi2017observation,kyprianidis2021observation,beatrez2023critical}.
\newline\indent
Beyond echo physics, related rotating-frame analogues have been exploited for audio-frequency control of nuclear spins in imaging~\cite{zhu2016selective} and for quantum sensing protocols that leverage long-lived prethermal order to transduce weak AC fields into a continuously measurable response~\cite{sahin2022high,harkins2024anomalously}. The prethermal rotating-frame solid echo reported here adds a new element to this toolkit: an unexpected refocusing channel within an interacting dipolar solid that can be used both as a diagnostic of Floquet micromotion and as a route to higher-throughput echo spectroscopy.

\emph{Acknowledgements.---}This work was funded by ONR (N00014-20-1-2806), AFOSR YIP, NSF MRI (2320520), AFOSR DURIP (FA9550-22-1-0156), and the CIFAR Azrieli Foundation (GS23-013).

\bibliographystyle{apsrev4-2}
\bibliography{plr_berkeley3.bib}

@article{abanin2015exponentially,
  title={Exponentially slow heating in periodically driven many-body systems},
  author={Abanin, Dmitry A. and {De Roeck}, Wojciech and Huveneers, Fran{\c{c}}ois},
  journal={Physical Review Letters},
  volume={115},
  pages={256803},
  year={2015},
  publisher={APS}
}

@article{ajoy2018enhanced,
  title={Enhanced dynamic nuclear polarization via swept microwave frequency combs},
  author={Ajoy, A and Nazaryan, R and Liu, K and Lv, X and Safvati, B and Wang, G and Druga, E and Reimer, JA and Suter, D and Ramanathan, C and others},
  journal={Proceedings of the National Academy of Sciences},
  volume={115},
  number={42},
  pages={10576--10581},
  year={2018},
  publisher={National Acad Sciences}
}

@article{ajoy2018orientation,
  title={Orientation-independent room temperature optical 13C hyperpolarization in powdered diamond},
  author={Ajoy, Ashok and Liu, Kristina and Nazaryan, Raffi and Lv, Xudong and Zangara, Pablo R and Safvati, Benjamin and Wang, Guoqing and Arnold, Daniel and Li, Grace and Lin, Arthur and others},
  journal={Science advances},
  volume={4},
  number={5},
  pages={eaar5492},
  year={2018},
  publisher={American Association for the Advancement of Science}
}

@article{ajoy2019hyperpolarized,
  title={Hyperpolarized relaxometry based nuclear T 1 noise spectroscopy in diamond},
  author={Ajoy, Ashok and Safvati, Ben and Nazaryan, Raffi and Oon, JT and Han, Ben and Raghavan, Priyanka and Nirodi, Ruhee and Aguilar, Alessandra and Liu, Kristina and Cai, Xiao and others},
  journal={Nature communications},
  volume={10},
  number={1},
  pages={5160},
  year={2019},
  publisher={Nature Publishing Group UK London}
}

@article{bain1978modulation,
  title={Modulation of NMR spin echoes in coupled systems},
  author={Bain, Alex D},
  journal={Chemical Physics Letters},
  volume={57},
  number={2},
  pages={281--284},
  year={1978},
  publisher={Elsevier}
}

@article{beatrez2021floquet,
  title={Floquet prethermalization with lifetime exceeding 90 s in a bulk hyperpolarized solid},
  author={Beatrez, William and Janes, Otto and Akkiraju, Amala and Pillai, Arjun and Oddo, Alexander and Reshetikhin, Paul and Druga, Emanuel and McAllister, Maxwell and Elo, Mark and Gilbert, Benjamin and others},
  journal={Physical review letters},
  volume={127},
  number={17},
  pages={170603},
  year={2021},
  publisher={APS}
}

@article{beatrez2023critical,
  title={Critical prethermal discrete time crystal created by two-frequency driving},
  author={Beatrez, William and Fleckenstein, Christoph and Pillai, Arjun and de Leon Sanchez, Erica and Akkiraju, Amala and Diaz Alcala, Jesus and Conti, Sophie and Reshetikhin, Paul and Druga, Emanuel and Bukov, Marin and others},
  journal={Nature Physics},
  volume={19},
  number={3},
  pages={407--413},
  year={2023},
  publisher={Nature Publishing Group UK London}
}

@article{blanes2009magnus,
  title={The Magnus expansion and some of its applications},
  author={Blanes, S. and Casas, F. and Oteo, J. and Ros, J.},
  journal={Physics Reports},
  volume={470},
  pages={151},
  year={2009},
  doi={10.1016/j.physrep.2008.11.001}
}

@article{bukov2015prethermal,
  title={Prethermal floquet steady states and instabilities in the periodically driven, weakly interacting bose-hubbard model},
  author={Bukov, Marin and Gopalakrishnan, Sarang and Knap, Michael and Demler, Eugene},
  journal={Physical review letters},
  volume={115},
  number={20},
  pages={205301},
  year={2015},
  publisher={APS}
}

@article{bukov2015universal,
  title={Universal high-frequency behavior of periodically driven systems: from dynamical stabilization to Floquet engineering},
  author={Bukov, Marin and {D'Alessio}, Luca and Polkovnikov, Anatoli},
  journal={Advances in Physics},
  volume={64},
  number={2},
  pages={139--226},
  year={2015}
}

@article{choi2017observation,
  title={Observation of discrete time-crystalline order in a disordered dipolar many-body system},
  author={Choi, S. and Choi, J. and Landig, R. and Kucsko, G. and Zhou, H. and Isoya, J. and Jelezko, F. and Onoda, S. and Sumiya, H. and Khemani, V. and Lukin, M. D. and others},
  journal={Nature},
  volume={543},
  pages={221},
  year={2017}
}

@article{dalessio2014long,
  title={Long-time behavior of isolated periodically driven interacting lattice systems},
  author={{D'Alessio}, Luca and Rigol, Marcos},
  journal={Physical Review X},
  volume={4},
  pages={041048},
  year={2014},
  publisher={APS}
}

@article{else2016floquet,
  title={Floquet time crystals},
  author={Else, Dominic V. and Bauer, Bela and Nayak, Chetan},
  journal={Physical Review Letters},
  volume={117},
  pages={090402},
  year={2016},
  publisher={APS}
}

@article{else2017prethermal,
  title={Prethermal phases of matter protected by time-translation symmetry},
  author={Else, Dominic V. and Bauer, Bela and Nayak, Chetan},
  journal={Physical Review X},
  volume={7},
  pages={011026},
  year={2017},
  publisher={APS}
}

@article{else2020discrete,
  title={Discrete time crystals},
  author={Else, Dominic V and Monroe, Christopher and Nayak, Chetan and Yao, Norman Y},
  journal={Annual Review of Condensed Matter Physics},
  volume={11},
  number={1},
  pages={467--499},
  year={2020},
  publisher={Annual Reviews}
}

@book{ernst1987principles,
  title={Principles of Nuclear Magnetic Resonance in One and Two Dimensions},
  author={Ernst, R. and Bodenhausen, G. and Wokaun, A.},
  publisher={Clarendon Press},
  address={Oxford},
  year={1987}
}

@book{freeman1998spin,
  title={Spin choreography},
  author={Freeman, Ray},
  year={1998},
  publisher={Oxford University Press Oxford}
}

@article{goldman2014periodically,
  title={Periodically driven quantum systems: effective Hamiltonians and engineered gauge fields},
  author={Goldman, Nathan and Dalibard, Jean},
  journal={Physical Review X},
  volume={4},
  pages={031027},
  year={2014},
  publisher={APS}
}

@article{gopalakrishnan2007quenching,
  title={Quenching and recoupling of echo modulations in NMR spectroscopy},
  author={Gopalakrishnan, Karthik and Aeby, Nicolas and Bodenhausen, Geoffrey},
  journal={ChemPhysChem},
  volume={8},
  number={12},
  pages={1791--1802},
  year={2007},
  publisher={Wiley Online Library}
}

@book{haeberlen1976high,
  title={High Resolution NMR in Solids: Selective Averaging},
  author={Haeberlen, Ulrich},
  publisher={Academic Press},
  address={New York},
  year={1976}
}

@article{harkins2024anomalously,
  title={Anomalously extended Floquet prethermal lifetimes and applications to long-time quantum sensing},
  author={Harkins, Kieren A and Selco, Cooper and Bengs, Christian and Marchiori, David and Moon, Leo Joon Il and Zhang, Zhuo-Rui and Yang, Aristotle and Singh, Angad and Druga, Emanuel and Song, Yi-Qiao and others},
  journal={arXiv preprint arXiv:2410.09028},
  year={2024}
}

@article{ho2023quantum,
  title={Quantum and classical Floquet prethermalization},
  author={Ho, Wen Wei and Mori, Takashi and Abanin, Dmitry A and Dalla Torre, Emanuele G},
  journal={Annals of Physics},
  volume={454},
  pages={169297},
  year={2023},
  publisher={Elsevier}
}

@article{kyprianidis2021observation,
  title={Observation of a prethermal discrete time crystal},
  author={Kyprianidis, A. and Machado, F. and Morong, W. and Becker, P. and Collins, K. S. and Else, D. V. and Feng, L. and Hess, P. W. and Nayak, C. and Pagano, G. and Yao, N. Y. and Monroe, C.},
  journal={Science},
  volume={372},
  pages={1192},
  year={2021},
  doi={10.1126/science.abg8102}
}

@article{lazarides2014equilibrium,
  title={Equilibrium states of generic quantum systems subject to periodic driving},
  author={Lazarides, Achilleas and Das, Arnab and Moessner, Roderich},
  journal={Physical Review E},
  volume={90},
  pages={012110},
  year={2014},
  publisher={APS}
}

@article{li2007generating,
  title={Generating unexpected spin echoes in dipolar solids with $\pi$ pulses},
  author={Li, Dale and Dementyev, AE and Dong, Yanqun and Ramos, RG and Barrett, SE},
  journal={Physical review letters},
  volume={98},
  number={19},
  pages={190401},
  year={2007},
  publisher={APS}
}

@article{linden2009quantum,
  title={Quantum mechanical evolution towards thermal equilibrium},
  author={Linden, Noah and Popescu, Sandu and Short, Anthony J and Winter, Andreas},
  journal={Physical Review E—Statistical, Nonlinear, and Soft Matter Physics},
  volume={79},
  number={6},
  pages={061103},
  year={2009},
  publisher={APS}
}

@article{machado2020longrange,
  title={Long-range prethermal phases of nonequilibrium matter},
  author={Machado, Francisco and Else, Dominic V. and Kahanamoku-Meyer, Gregory D. and Nayak, Chetan and Yao, Norman Y.},
  journal={Physical Review X},
  volume={10},
  pages={011043},
  year={2020},
  publisher={APS}
}

@article{magnus1954exponential,
  title={On the exponential solution of differential equations for a linear operator},
  author={Magnus, Wilhelm},
  journal={Communications on Pure and Applied Mathematics},
  volume={7},
  pages={649},
  year={1954}
}

@article{mananga2011introduction,
  title={Introduction of the Floquet-Magnus expansion in solid-state nuclear magnetic resonance spectroscopy},
  author={Mananga, Eugene S and Charpentier, Thibault},
  journal={The Journal of chemical physics},
  volume={135},
  number={4},
  year={2011},
  publisher={AIP Publishing}
}

@article{mansfield1965multiple,
  title={Multiple-pulse nuclear magnetic resonance transients in solids},
  author={Mansfield, P},
  journal={Physical Review},
  volume={137},
  number={3A},
  pages={A961},
  year={1965},
  publisher={APS}
}

@article{mansfield1971pulsed,
  title={Pulsed NMR in solids},
  author={Mansfield, Prog},
  journal={Progress in nuclear magnetic resonance spectroscopy},
  volume={8},
  number={1},
  pages={41--101},
  year={1971},
  publisher={Elsevier}
}

@article{maricq1987spin,
  title={Spin thermodynamics of periodically time-dependent systems: The quasistationary state and its decay},
  author={Maricq, M Matti},
  journal={Physical Review B},
  volume={36},
  number={1},
  pages={516},
  year={1987},
  publisher={APS}
}

@article{mori2016rigorous,
  title={Rigorous bound on energy absorption and generic relaxation in periodically driven quantum systems},
  author={Mori, Takashi and Kuwahara, Takashi and Saito, Keiji},
  journal={Physical Review Letters},
  volume={116},
  pages={120401},
  year={2016},
  publisher={APS}
}

@article{mori2018thermalization,
  title={Thermalization and prethermalization in isolated quantum systems: a theoretical overview},
  author={Mori, Takashi and Ikeda, Tatsuhiko N and Kaminishi, Eriko and Ueda, Masahito},
  journal={Journal of Physics B: Atomic, Molecular and Optical Physics},
  volume={51},
  number={11},
  pages={112001},
  year={2018},
  publisher={IOP Publishing}
}

@article{peng2021floquet,
  title={Floquet prethermalization in dipolar spin chains},
  author={Peng, P. and Yin, C. and Huang, X. and Ramanathan, C. and Cappellaro, P.},
  journal={Nature Physics},
  volume={17},
  pages={444},
  year={2021}
}

@article{pillai2023electron,
  title={Electron-to-nuclear spectral mapping via dynamic nuclear polarization},
  author={Pillai, Arjun and Elanchezhian, Moniish and Virtanen, Teemu and Conti, Sophie and Ajoy, Ashok},
  journal={The Journal of Chemical Physics},
  volume={159},
  number={15},
  year={2023},
  publisher={AIP Publishing}
}

@article{powles1962double,
  title={Double-pulse nuclear-resonance transients in solids},
  author={Powles, JG and Mansfield, Peter},
  journal={Physics Letters},
  volume={2},
  number={2},
  pages={58--59},
  year={1962}
}

@article{powles1962measurement,
  title={Measurement of J coupling using spin echo techniques},
  author={Powles, JG and Strange, JH},
  journal={Discussions of the Faraday Society},
  volume={34},
  pages={30--37},
  year={1962},
  publisher={Royal Society of Chemistry}
}

@article{ramanathan2024universal,
  title={Universal dynamics exposed by interaction quenches},
  author={Ramanathan, Chandrasekhar},
  journal={Nature Physics},
  pages={1--2},
  year={2024},
  publisher={Nature Publishing Group UK London}
}

@article{rhim1976multiple,
  title={Multiple-pulse spin locking in dipolar solids},
  author={Rhim, W-K and Burum, DP and Elleman, DD},
  journal={Physical Review Letters},
  volume={37},
  number={26},
  pages={1764},
  year={1976},
  publisher={APS}
}

@article{rubio2020floquet,
  title={Floquet prethermalization in a bose-hubbard system},
  author={Rubio-Abadal, Antonio and Ippoliti, Matteo and Hollerith, Simon and Wei, David and Rui, Jun and Sondhi, SL and Khemani, Vedika and Gross, Christian and Bloch, Immanuel},
  journal={Physical Review X},
  volume={10},
  number={2},
  pages={021044},
  year={2020},
  publisher={APS}
}

@article{saha2023cascaded,
  title={Cascaded dynamics of a periodically driven dissipative dipolar system},
  author={Saha, Saptarshi and Bhattacharyya, Rangeet},
  journal={Physical Review A},
  volume={107},
  number={2},
  pages={022206},
  year={2023},
  publisher={APS}
}

@article{sahin2022continuously,
  title={Continuously tracked, stable, large excursion trajectories of dipolar coupled nuclear spins},
  author={Sahin, Ozgur and Asadi, Hawraa Al and Schindler, Paul and Pillai, Arjun and Sanchez, Erica and Markham, Matthew and Elo, Mark and McAllister, Maxwell and Druga, Emanuel and Fleckenstein, Christoph and others},
  journal={arXiv preprint arXiv:2206.14945},
  year={2022}
}

@article{sahin2022high,
  title={High field magnetometry with hyperpolarized nuclear spins},
  author={Sahin, Ozgur and de Leon Sanchez, Erica and Conti, Sophie and Akkiraju, Amala and Reshetikhin, Paul and Druga, Emanuel and Aggarwal, Aakriti and Gilbert, Benjamin and Bhave, Sunil and Ajoy, Ashok},
  journal={Nature communications},
  volume={13},
  number={1},
  pages={5486},
  year={2022},
  publisher={Nature Publishing Group UK London}
}

@article{solomon1958multiple,
  title={Multiple echoes in solids},
  author={Solomon, I},
  journal={Physical Review},
  volume={110},
  number={1},
  pages={61},
  year={1958},
  publisher={APS}
}

@article{vega1985relaxation,
  title={Relaxation in spin-echo and spin-lock experiments},
  author={Vega, Alexander J},
  journal={Journal of Magnetic Resonance (1969)},
  volume={65},
  number={2},
  pages={252--267},
  year={1985},
  publisher={Elsevier}
}

@article{waugh1967multiple,
  title={Multiple spin echoes in dipolar solids},
  author={Waugh, JS and Wang, CH},
  journal={Physical Review},
  volume={162},
  number={2},
  pages={209},
  year={1967},
  publisher={APS}
}

@article{weidinger2017floquet,
  title={Floquet prethermalization and regimes of heating in a periodically driven, interacting quantum system},
  author={Weidinger, Simon A and Knap, Michael},
  journal={Scientific reports},
  volume={7},
  number={1},
  pages={45382},
  year={2017},
  publisher={Nature Publishing Group UK London}
}

@article{wells1965nmr,
  title={NMR spin-echo trains for a coupled two-spin system},
  author={Wells, EJ and Gutowsky, HS},
  journal={The Journal of Chemical Physics},
  volume={43},
  number={9},
  pages={3414},
  year={1965}
}

@article{wilcox1967exponential,
  title={Exponential operators and parameter differentiation in quantum physics},
  author={Wilcox, R. M.},
  journal={Journal of Mathematical Physics},
  volume={8},
  pages={962},
  year={1967}
}

@article{zhu2016selective,
  title={Selective magnetic resonance imaging of magnetic nanoparticles by acoustically induced rotary saturation},
  author={Zhu, Bo and Witzel, Thomas and Jiang, Shan and Huang, Susie Y and Rosen, Bruce R and Wald, Lawrence L},
  journal={Magnetic resonance in medicine},
  volume={75},
  number={1},
  pages={97--106},
  year={2016},
  publisher={Wiley Online Library}
}

@incollection{maricq1990long,
  title={Long-time limitations of the average hamiltonian theory: A dressed-states viewpoint},
  author={Maricq, M Matti},
  booktitle={Advances in Magnetic and Optical Resonance},
  volume={14},
  pages={151--182},
  year={1990},
  publisher={Elsevier}
}

@article{sakellariou1998quasi,
  title={Quasi equilibria in solid-state NMR},
  author={Sakellariou, Dimitris and Hodgkinson, Paul and Emsley, Lyndon},
  journal={Chemical physics letters},
  volume={293},
  number={1-2},
  pages={110--118},
  year={1998},
  publisher={Elsevier}
}

@article{moon2025high,
  title={High-speed, high-memory NMR spectrometer and hyperpolarizer},
  author={Moon, Leo Joon Il and Beatrez, William and Ball, Jason and Mercade, Joan and Elo, Mark and Singh, Angad and Druga, Emanuel and Ajoy, Ashok},
  journal={Journal of Magnetic Resonance},
  pages={107952},
  year={2025},
  publisher={Elsevier}
}

@article{henrichs1984nuclear,
  title={Nuclear spin-lattice relaxation via paramagnetic centers in solids. 13C NMR of diamonds},
  author={Henrichs, P Mark and Cofield, Milton L and Young, Ralph H and Hewitt, J Michael},
  journal={Journal of Magnetic Resonance (1969)},
  volume={58},
  number={1},
  pages={85--94},
  year={1984},
  publisher={Elsevier}
}

\end{document}